\documentclass[fleqn,usenatbib]{mnras}

\usepackage{mathptmx}
\usepackage[T1]{fontenc}
\usepackage{ae,aecompl}

\usepackage{graphicx}	% Including figure files
\usepackage{amsmath}	% Advanced maths commands
\usepackage{amssymb}	% Extra maths symbols

\let\la=\lesssim
\title[The magnetic Rayleigh-Taylor instability]{On the nature of the magnetic Rayleigh-Taylor instability in Astrophysical Plasma: The case of uniform magnetic field strength}

\author[A. S. Hillier]{
Andrew S. Hillier$^{1}$\thanks{E-mail: ah826@cam.ac.uk}
\\
$^{1}$Department of Applied Mathematics and Theoretical Physics, University of Cambridge, Wiberforce Road,  Cambridge CB3 0WA, UK\\
}

\date{Accepted XXX. Received YYY; in original form ZZZ}

\pubyear{2015}

\begin{document}
\label{firstpage}
\pagerange{\pageref{firstpage}--\pageref{lastpage}}
\maketitle

\begin{abstract}
The magnetic Rayleigh-Taylor instability has been shown to play a key role in many astrophysical systems.
The equation for the growth rate of this instability in the incompressible limit, and the most-unstable mode that can be derived from it, are often used to estimate the strength of the magnetic field that is associated with the observed dynamics.
However, there are some issues with the interpretations given.
Here we show that the class of most unstable modes $k_u$ for a given $\theta$, the class of modes often used to estimate the strength of the magnetic field from observations, for the system leads to the instability growing as $\sigma^2=1/2 A g k_u$, a growth rate which is independent of the strength of the magnetic field and which highlights that small scales are preferred by the system, but not does not give the fastest growing mode for that given $k$.
We also highlight that outside of the interchange ($\mathbfit{k}\cdot\mathbfit{B}=0$) and undular ($\mathbfit{k}$ parallel to $\mathbfit{B}$) modes, all the other modes have a perturbation pair of the same wavenumber and growth rate that when excited in the linear regime {can result} in an interference pattern that gives field aligned filamentary structure often seen in 3D simulations.
{The analysis was extended to a sheared magnetic field, where it was found that it was possible to extend the results for a non-sheared field to this case. We suggest that without magnetic shear it is too simplistic to be used to infer magnetic field strengths in astrophysical systems.}
\end{abstract}


%%%%%%%%%%%%%%%%%%%%%%%%%%%%%%%%%%%%%%%%%%%%%%%%%%

%%%%%%%%%%%%%%%%% BODY OF PAPER %%%%%%%%%%%%%%%%%%

\section{The Rayleigh-Taylor Instability in Astrophysical Plasma}\label{INTRO}

The Rayleigh-Taylor instability, first proposed by \cite{RAY1900} and \cite{TAY1950}, is a fundamental process in many space and astrophysical systems.
For a contact discontinuity that is formed where a heavy fluid is supported above a light fluid against gravity, this boundary is unstable to perturbations that grow by converting gravitational potential energy into kinetic energy creating rising and falling fingers.
The growth rate ($\sigma$) for this instability in the absence of magnetic field or viscous effects is give as:
\begin{equation}\label{RThydro}
\sigma=\sqrt{A k g},
\end{equation}
where $k$ is the wavenumber, $g$ is the acceleration due to gravity and A is the Atwood number give as $(\rho_u - \rho_l)/(\rho_u + \rho_l)$ for where $\rho_u$ is the upper density and $\rho_l$ is the lower density \citep{CHAN1961}.
\citet{DALY1967} described how the evolution of the Rayleigh-Taylor instability changes from the symmetry of the low Atwood number Rayleigh-Taylor instability to be replaced by the formation of rising bubbles and sharp falling spikes in the high Atwood number limit.
For a review of the hydrodynamic Rayleigh-Taylor instability see, for example, \cite{Sharp1984}.

The inclusion of a horizontal magnetic field to the Rayleigh-Taylor instability adds a directionality to the system \citep{KRUS1954}.
The interchange mode, where the wavevector $\mathbfit{k}$ is perpendicular to the magnetic field and so magnetic tension does not have an effect, reduces the problem to one analogous to the hydrodynamic situation where a total pressure replaces the role of the gas pressure.
The undular mode, where the wavevector $\mathbfit{k}$ is parallel to the magnetic field and as such drives distortion of the magnetic field, creates a magnetic tension force that works to suppress high wavenumber perturbations along the magnetic field.
The growth rate of the magnetic Rayleigh-Taylor instability for a mixed mode perturbation where the magnetic field is only in the $y$-direction is given as
\begin{equation}\label{RTeqn}
\sigma^2=kg\left[A- \frac{B^2k_y^2}{2 \pi (\rho_u + \rho _l)gk } \right ],
\end{equation}
where $B$ is the magnetic field strength in the $y$ direction \citep{CHAN1961}.
This implies that the system is always unstable providing a perturbation with sufficiently small $k_x$ is given.
For Equation \ref{RTeqn} a critical wavelength ($\lambda_c$) that gives a growth rate of $\sigma=0$ can be defined as $\lambda_c=2\pi/k_c=B^2cos^2\theta/(\rho_u-\rho_l)g$, where $\theta$ is the angle between the $k$ vector and the magnetic field.

The Rayleigh-Taylor instability drives many observed features in astrophysical systems.
\citet{HAC1989} described how the instability can lead to element mixing in supernova explosions.
\citet{HEST1996} compared the observational characteristics of the Crab Nebula with simulations of the magnetic Rayleigh-Taylor instability performed by \citet{JUN1995}, finding that the magnetic Rayleigh-Taylor instability could explain the observed filamentary structure. 
{Recent axisymmetric simulations of the Crab Nebula by \citet{PORTH2014}, using adaptive mesh refinement to provide high resolution, found that the magnetic field is insufficient to  suppress the growth of the instability.}
In the Earth's ionosphere, the rise of regions of depleted plasma against the gravitational field during the Equatorial Spread-F phenomenon has been interpreted as the occurrence of the magnetic Rayleigh-Taylor instability in a low-beta magnetic plasma environment \citep{Kelley1976,Taka2009}.
{\citet{GRAT1996} described how variations in the solar wind can lead to expansion and contraction of the magnetopause pointing out that in the expansion phase there is an effective gravity that drives the magnetic Rayleigh-Taylor instability, a similar physical process can also happen at the heliopause as a result of changes due to the solar cycle \citep{BORO2014}.}
Observations by \citet{BER2008, BER2010} show this instability occurring in quiescent prominences. 
This instability has also been found in plasma jet experiments \citep{MOS2012}

Numerical investigations in the magnetic Rayleigh-Taylor instability in 3D have revealed a great deal about its evolution.
One of the key aspects that these simulations have revealed is that in the 3D magnetic Rayleigh-Taylor simulations is that in 3D the instability results in the formation of structures that are elongated in the direction of the magnetic field \citep{STONE2007}.
This feature has been of particular importance in understanding the formation of plumes in solar prominences \citep{HILL2011,HILL2012, TERR2015, KEP2015} or in the formation of filamentary structure associated with emerging magnetic flux \citep{IS2006}.

{To make the linear theory for the magnetic Rayleigh-Taylor instability more applicable to the astrophysical settings described above, there has been development of the linear theory beyond the simple model used by \citet{CHAN1961}.
\citet{RUD2014} investigated the role of magnetic shear on the instability both for the single discontinuity and for a dense slab embedded in a tenuous atmosphere.
This can be treated as an extension of case of a plasma interacting with a vacuum where there is shear in the magnetic field as elucidated in Chapter 6 of \citet{GOED2004}, their Equation 6.612 in the short wavelength limit (when compared to distance to the upper and lower walls) matches that of Equation 19 of \citet{RUD2014} where $\rho_e$ is taken as $0$.
One key role of the magnetic shear was to bound the growth of the instability, i.e. give the highest growth rate for a finite $k$, as the pure interchange mode was no longer possible in the system.
\citet{LIB2008} and \citet{LIB2009} investigated the role of compressibility and stratification on the growth of the instability in an isothermal plasma.
They found that the role of stratification weakened the instability but that compressibility has a destabilizing effect.
{As in many astrophysical systems} the temperatures are insufficient to fully ionise the plasma resulting in partially ionised plasma, \citet{DIAZ2012} investigated how the ion-neutral interaction through collisions changes the growthrate in the compressible magnetic Rayleigh-Taylor instability. 
Their analysis showed {that though} these physics do not alter the threshold for instability, the ion-neutral collisions can vastly reduce the growth rate of the instability.
\citet{ZHAI2016} developed a model of a hybrid of the current-driven kink instability and the magnetic Rayleigh-Taylor instability in a flux tube to model the experimental findings of \citet{MOS2012}, where they found this new geometry was able to explain the growth of the observed instability.
For a detail investigation into many of the important process relating to the Rayleigh-Taylor instability including the extension to stratified atmospheres, compressibility, inhomogeneity and continua see, for example, \citet{GOED2004} and \citet{GOED2010}.}

One of the key application of equation for the linear growth rate of the instability has been in estimation of the magnetic field strength of various observed phenomena.
From observations, the wavelength of the magnetic Rayleigh-Taylor instability is determined and, assuming that this wavelength is created by the most unstable mode, the observed wavelength can be connected with the magnetic field strength through the following equation:
\begin{equation}\label{UNST}
k_u=\frac{\pi g(\rho_u-\rho_l)}{ \cos^2 \theta B^2},
\end{equation}
where $\cos \theta$ is the angle between the wavenumber $k$ and the magnetic field direction.
This has been applied to determine the magnetic field strength in a wide variety of situations including prominence plumes \citep{RYU2010}, material from a solar eruption \citep{CAR2014,INN2012} and for supernova remnants \citep{HEST1996}.

In this paper, we revisit the analysis of the magnetic Rayleigh-Taylor instability to understand what will happen to a system in the linear stages of its evolution when given a random perturbation and thus exciting the most unstable modes of the system.
{We first look at the case of a uniform magnetic field and then extend these results to include the influence of magnetic shear at the boundary.}

\section{What are the critical and most unstable modes?}

\subsection{Some insight from Hydrodynamics}

Following \citet{CHAN1961} it is possible to show that the hydrodynamic Rayleigh-Taylor instability with surface tension acting on the boundary between the light and dense fluids follows the following growth rate:
\begin{equation}\label{hydro_tension}
\sigma^2=gk\left[A - \frac{k^2 T}{g(\rho_u+\rho_l)} \right],
\end{equation}
where $T$ is the surface tension.
By taking the derivative of this equation with respect to $k$ and looking for solutions where $\partial \sigma/\partial k =0$ gives a most unstable mode $k_u$ of:
\begin{equation}\label{hydro_tension_length}
k_u=\left[ \frac{g(\rho_u-\rho_l) }{3 T} \right]^{1/2}
\end{equation}
which bears some similarity to Equation \ref{UNST}.

It is important to note one crucial point here. The growth rate in Equation \ref{hydro_tension} describes an isotropic system in the $x$--$y$ plane, but Equation \ref{RTeqn} is not isotropic as the inclusion of the magnetic field gives a dependence on $\theta$ to the growth rate.
Therefore it is important to understand that you can define a most unstable mode for a fixed $k$ or one for a fixed $\theta$ (the angle between $\mathbfit{k}$ and $\mathbfit{B}$), the latter is that which was used for Equation \ref{UNST}.

\subsection{The most unstable mode for a given $\theta$}

Equation \ref{UNST} gives the most unstable mode, but what does this actually imply for the development of the instability?
To look at this, we only need the equation for the growth rate of the RT:
\begin{equation}
\sigma^2=Akg-\frac{k_y^2B^2}{2\pi(\rho_u+\rho_l)},
\end{equation}
If we first perform a thought experiment, we can understand a very simple property of this equation, i.e. that the scale across the magnetic field determines what scale can form along the magnetic field.
If we start with $\mathbfit{k}=[0,k_y]$ we have only an undular mode, which has a critical wavenumber $k_{crit}$ such that $\sigma=0$.
If we now increase $k$ from this $k_{crit}$ by only increasing $k_x$ the first term on the RHS of the equation increases, but the second term stays the same, making $\sigma >0$ so the system is unstable.
Therefore, to return to the critical wavelength for this now non-zero $k_x$, the wavenumber along the magnetic field has to be increased.
This means that the smaller the scale we have across the field the smaller the scale that can form along the field.
You can understand this physically by comparing the gravitational freefall time to the timescale for magnetic suppression. 
The critical mode is when these are balanced, so by making the freefall time smaller by reducing the lengthscale across the magnetic field, to maintain marginal stability the timescale for magnetic suppression has to go down by making that lengthscale smaller as well.
It is actually quite trivial to show that if we set $k_x$ we can find a $k_y$ s.t. we find the most unstable mode for that $k_x$.

We can define the equation for the critical wavenumber as:
\begin{equation}
0=A g \sqrt{k_x^2 + k_y^2}-\frac{k_y^2 B^2}{2 \pi (\rho_u  +\rho_l)},
\end{equation}
where $k_y$ is the part of the wavenumber parallel to the magnetic field and $k_x$ is the part of the wavenumber perpendicular to the magnetic field.
This can be rearranged to give $k_x$ as a function of $k_y$:
\begin{align}
k_x^2 = & \left(\frac{k_y^2 B^2}{2 \pi g (\rho_u - \rho_l)}\right)^2-k_y^2 \\
      = & k_y^2\left[ k_y^2\left(\frac{ B^2}{2 \pi g (\rho_u - \rho_l)}\right)^2 -1 \right].
\end{align}
Using this equation we can now investigate the relation between $k_x$ and $k_y$ in terms of the stability of the system.

Firstly, the simplest approach is to investigate the roots of the equation.
One of these roots exists at $k_y=0$.
The other can be found by solving
\begin{equation}
 k_y^2\left(\frac{ B^2}{2 \pi g (\rho_u - \rho_l)}\right)^2 =1
\end{equation}
which gives
\begin{equation}\label{kxmax}
 k_y=\frac{2 \pi g (\rho_u - \rho_l)}{ B^2}.
\end{equation}
This region of $k_y$ perturbations gives the region in which pure undular modes can exist, i.e. even if $k_x$ is 0 a perturbation can grow.

\begin{figure*}[ht]
  \begin{center}
\includegraphics[width=14cm]{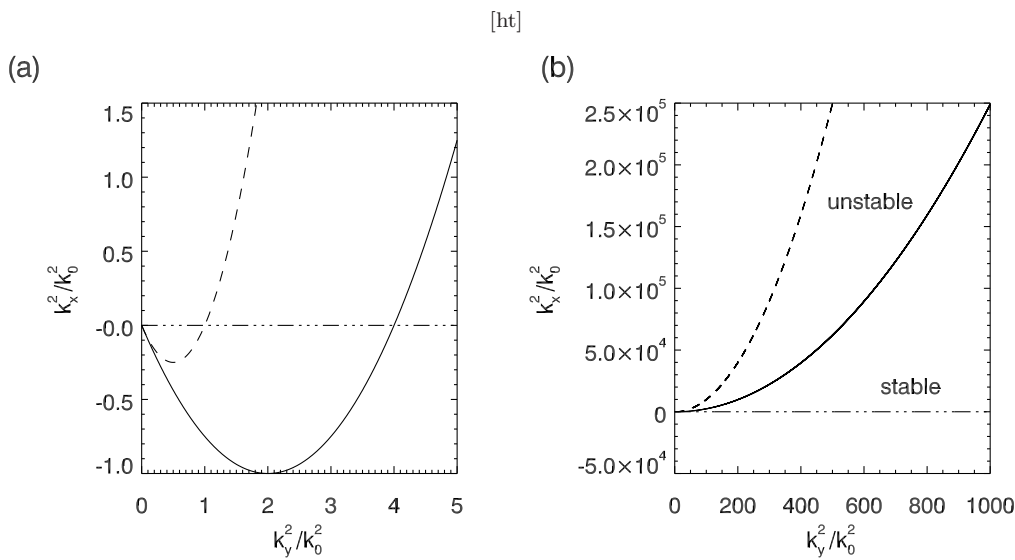}

  \end{center}
  \caption{ The relation of critical (solid line) and most unstable mode (dashed line) of a given perturbation across the magnetic field. Here the normalising lengthscale $k_0$ is set as the most unstable undular mode. The triple-dot-dashed line marks the boundary between the physically realisable perturbations and those that are not, i.e. $k_x^2<0$.}
\label{unstable_mode}
\end{figure*}

Figure \ref{unstable_mode} shows the relation between different $k_x$ and $k_y$ values where panel (a) shows a small region in k-space that highlights the $0$ crossing of $k_y^2$ as part of the distribution shown in panel (b).
The solid line denotes the most critical mode.
In panel (b), the regions of stability and instability are labelled.
The region below the dashed line in panel (a) gives the values for $k_y$ where a pure undular mode can be excited.
For larger $k_y$ values, it is not possible to excite a pure undular mode, and so a mixed mode (where $k_x>0$) is required.
Therefore, to excite a mode along the magnetic field greater than the value for $k_y$ given in Equation \ref{kxmax}, a mode including a component perpendicular to the magnetic field must be excited.

We can perform the same analysis for the most unstable growth rate by differentiating by $k$:
\begin{align}
k_x^2= & \left(\frac{k_y^2 B^2}{\pi g (\rho_u - \rho_l)}\right)^2-k_y^2\\
= & k_y^2\left[ k_y^2\left(\frac{B^2}{\pi g (\rho_u - \rho_l)}\right)^2 -1 \right]\nonumber\\
\implies k_y^2 = & \frac{1}{2}\left(\frac{ \pi g (\rho_u - \rho_l)}{ B^2}\right)^2\label{kx_ky_dep}\\&+\frac{1}{2}\left(\frac{ \pi g (\rho_u - \rho_l)}{ B^2}\right)\left[\left(\frac{ \pi g (\rho_u - \rho_l)}{ B^2}\right)^2 + 4k_x^2 \right]^{1/2}. \nonumber
\end{align}
Note that the second solution of the negative of the square root cannot lead to any physically realisable solutions and has been neglected.
Using this equation we can now investigate the relation between $k_x$ and $k_y$ in terms of the most unstable mode of the system for a given angle $\theta$ between the magnetic field and the perturbation.

Before looking into the solutions of the equation, it is important to understand what exactly we mean by the most unstable mode.
Returning to the equation for the growth rate
\begin{equation}
\sigma^2=A g \sqrt{k_x^2 + k_y^2}-\frac{k_y^2 B^2}{2 \pi (\rho_u  +\rho_l)}
\end{equation}
if we set $k_y$ to be constant and only vary $k_x$ we can see that the larger $k_x$, the larger $\sigma$ becomes.
Therefore, there is no most unstable mode for a fixed $k_y$, as $k \rightarrow \infty$ so does $\sigma$.
This implies that when we talk about the most unstable $k$ for a given $\theta$, this also means we are discussing the most unstable $k_y$ in terms of a constant $k_x$, i.e. the preferential $k_y$ as associated with a $k_x$, as is shown in Equation \ref{kx_ky_dep}.

The dashed line in Figure \ref{unstable_mode} shows the relationship between $k_x$ and $k_y$ for the most unstable mode.
The point where this crosses the x-axis gives the most unstable undular mode.
It is clear that the most unstable modes that exist for higher $k_y$ are associated higher and higher values of $k_x$.
Therefore, the larger the $k_x$ value the larger the corresponding $k_y$ for the most unstable mode.
For then case where:
\begin{equation}
4k_x^2 \gg \left(\frac{ \pi g (\rho_u - \rho_l)}{ B^2}\right)^2
\end{equation}
then Equation \ref{kx_ky_dep} tells us that $k_y^2 \propto k_x$
For this, it can be concluded that for the very specific system that is under study, as the value of $k_x \rightarrow \infty$ then $k_y \rightarrow \infty$ but $k_y/k_x \rightarrow \infty$.

Another way to look at this system is to calculate the growth rate associated with each of the most unstable modes.
Firstly, we need to calculate the balance of the gravitational and magnetic terms for the most unstable mode, which are given by:
\begin{equation}
\frac{k_u cos^2\theta B^2}{2\pi(\rho_u + \rho_l)}=\frac{1}{2}Ag.
\end{equation}
This can be substituted into the equation for the growth rate, i.e.:
\begin{equation}
\sigma^2=A k_ug - \frac{k_u^2 cos^2\theta B^2}{2 \pi (\rho_u  +\rho_l)},
\end{equation}
where we have used the most unstable mode in this case.
This then gives:A
\begin{equation}\label{mum}
\sigma^2=\frac{1}{2}A g k_{u}.
\end{equation}
This equation is very interesting for reason that it does not depend on the magnetic field strength (though this is inherent in the aspect ratio that makes up the wavenumber $k_u$).
Direct comparison to the purely hydrodynamic case (see equation \ref{RThydro}) highlights that the addition of a magnetic field, regardless of strength of that magnetic field, reduces the growth rate by a factor of $1/\sqrt{2}$.
 point that should be taken from this is that there is no preferred lengthscale, simply put the smaller (i.e. the larger $k$) the better.

\subsection{The most unstable mode for a given $k$}

It is easy to calculate the most unstable mode for a given $k$ by taking the $\theta$ derivative of Equation \ref{RTeqn} and looking for the $0$ crossings of that function.
This gives:
\begin{equation}
\sin(2\theta)=0
\end{equation}
implying that $\theta=0,\pm \pi/2,\pm \pi ...$.
A maximum of this can be found when $\theta=\pi/2$, i.e. when $\mathbfit{k}\cdot\mathbfit{B}=0$.
Therefore, the most unstable mode for a given k is simply when the interchange mode is evoked:
\begin{equation}
\sigma^2=Akg.
\end{equation}

It is clear from 3D simulations of the magnetic Rayleigh-Taylor instability that modes with both a component perpendicular and parallel to the magnetic field are favoured \citep[e.g.][]{STONE2007}.
Given that the interchange mode has the largest growth rate of any mode of the same $k$, it is necessary to understand the why simulations give something different.
{It is necessary to note here that the work of \citet{STONE2007} used a compressible code to study the instability in a regime where the compressibility of the modes was estimated to be negligible.}

One other point that should be made is that for the most unstable mode for a given $k$ we find $\sigma^2=Akg$ and for a given $\theta$ we find $\sigma^2=1/2Akg$.
It turns out that for any constant $C \in \rm I\!R$ and $C \le 1$ a class of modes can be defined s.t. $\sigma^2=CAkg$.
For all $C$ the minimum $k$ for which this holds is given by:
\begin{equation}
k=k_y=2\pi(1-C)\frac{Ag(\rho_u+\rho_l)}{B^2}
\end{equation}
and the maximum is $k=\infty$.

\section{The principle of superposition and why 3D RT simulations give structure aligned with the magnetic field }\label{BASIS}

In this section I extend the statements of the previous section to explain why 3D simulations of the magnetic Rayleigh-Taylor instability can create structure aligned with the direction of the magnetic field, c.f. \citet{STONE2007} for the classic instability situation or, for example, \citet{HILL2012,IS2006} for specific simulations of astrophysical phenomena.
Here two important properties of the instability are used: the principle of superposition and the duality of the modes of the instability.
The first means that if you apply two linear perturbations to a system simultaneously, there is no coupling between the two modes.
The second of these refers to the fact that there are 2 modes, which form a set of basis vectors in $\rm I\!R^2$, of the magnetic RT that always have the same growth rate and wavenumber $k$, i.e. the wavenumber associated with the wavevectors $\mathbfit{k}=[k_x,k_y]$ and $\mathbfit{k}=[-k_x,k_y]$ (note that $\mathbfit{k}=[k_x,-k_y]$ and $\mathbfit{k}=[-k_x,-k_y]$ are actually the same pair).

\begin{figure*}[ht]
  \begin{center}
\includegraphics[width=16cm]{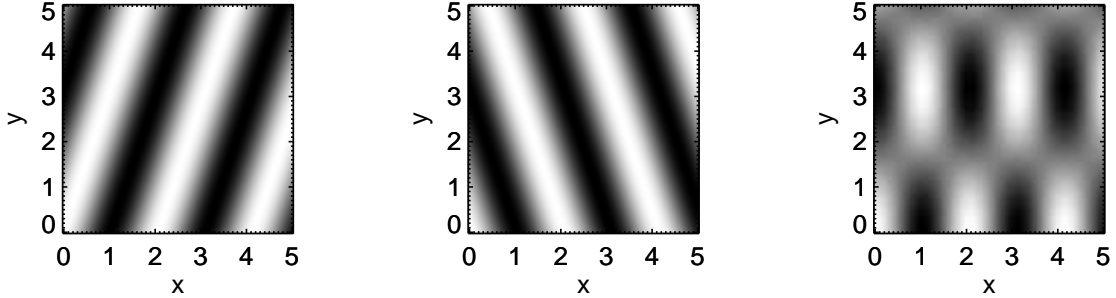}
  \end{center}
  \caption{Superposition of a pair of perturbations to give field aligned (in y-direction) structure formation.}
\label{superposition}
\end{figure*}

This is important to distinguishes the pure undular and interchange modes from all the other perturbations.
This happens because the pair of modes form a set of basis vectors for $\rm I\!R^2$ if the perturbation is not either a pure undular or interchange mode, i.e. to have a perturbation for which there exists another wave vector of same magnitude $k$, strictly speaking the condition is that for both perturbations $|k_x|$ and $|k_y|$ are the same, that has the same growth rate, then the perturbation must be a mixed mode.
The interchange (or undular) mode, however, becomes $\mathbfit{k}=[k_x,0]$ and $\mathbfit{k}=[-k_x,0]$ that are fundamentally the same perturbation to the system (i.e. does not form a set of basis vectors).
Figure \ref{superposition} shows how a pair of perturbations, $\mathbfit{k}=[3,1]$ and $\mathbfit{k}=[-3,1]$, can be superimposed to give structure that is longer in the y-direction than the x and the values of $\mathbfit{k}$ used ($3$ and $1$) tell you the wavenumber of the structure size in each direction.
This is very simple to think about for the hydrodynamic limit of this instability.
In this case the wavevectors $\mathbfit{k}=[k_x,k_Y]$ and $\mathbfit{k}=[-k_x,k_y]$ have the same growthrate when $k_x=k_y$, which creates creates rising and falling axially symmetric plumes of material.

We must note that this is inherently different from a superposition of an undular and interchange mode.
For example, if we take an undular mode of $k_y=1$ and an interchange mode of $k_x=3$, then there superposition would give the require structure, but the growth rates for the two perturbations would be very different, meaning that one perturbation had grown and reached its nonlinear saturation before the other had even started, i.e. they wouldn't grow coherently therefore not giving the growing filamentary structure.
{One caveat to the superposition argument is that for this to work exactly in a linear system, not only does the mode have to have the same wavenumber and growthrate, but also start growing at the same time and have the same amplitude to be completely coherent and create an interference pattern.
As this interference pattern is a clear signature of simulations, greater investigation of 3D numerical simulations is necessary to determine how these two Fourier components can start growing together.}

An interesting comparison can be made here with Faraday waves \citep{FARA1831}, which ties in very strongly to the similarity between the magnetic Rayleigh-Taylor and the hydrodynamic case with surface tension.
In this case the structure that forms is as a result of the resonant interaction of the excited waves.
For two wave interactions where $k_1=k_2$, i.e. the wavenumber is the same for each perturbation, formation of squares or rectangles is common, which matches very closely with my predicted structure formation for two perturbations of the same $k$ and $\sigma$ from the linear analysis (see Figure \ref{superposition}).
For Faraday waves, where there are two or more driving frequencies, where the frequencies form an integer ratio, the selection of different modes from each frequency can result in stark differences in the nonlinear structure formed \citet{TAKA2011}.
A weakly nonlinear analysis of the magnetic Rayleigh-Taylor instability for the simultaneous excitation of a number of system may lead to some very interesting results.

\subsection{Applicability of incompressible modes to simulations using compressible codes}

{From linear analysis, it is clear that if the exact eigenfunction of the linear mode is used to seed the instability in a numerical simulation, then the seeded mode should grow with the expected growth rate.
{This leads to the} question as to why simulations of the magnetic Rayleigh-Taylor instability are not completely dominated by interchange modes, which are the fastest growing modes.

There are a number of issues that may result in numerical simulations not preferentially picking the interchange modes, the first relates to the applicability of the approximations used to the numerical setup and scheme.
Here it is worth noting that though in the linear theory the weak magnetic field limit does not tend towards the hydrodynamic regime, where isotropic structures dominate, but simulations, e.g. \citet{STONE2007}, suggest the weak filed limit is tending towards the hydrodynamic limit.
The use of a compressible scheme is one point that could effect the modes that grow.
However, even for a compressible code, the mode will well approximate the incompressible limit if $\sqrt{g\lambda}/C_s\ll 1$, where $\lambda$ is the wavelength of the mode.
Evidence for this can be seen in \citet{LIB2008}, where high $k$ (small $\lambda$) modes match the incompressible limit even though compressibility is included.
The finite width over which the density varies and the numerical diffusion/viscosity of the scheme used may also play a role in determining the most unstable mode.
However, none of these considerations remove the simple fact that if the magnetic field is bent it will work to straighten itself, reducing the growth of those modes compared to a pure interchange mode of the same wavenumber $k$.
Therefore, the fastest growing mode being an interchange mode cannot be avoided.

Another possible explanation is that the timescale for the formation or the eigenfunction has some dependence on the Fourier mode being excited.
As most simulations will excite the instability by a random perturbation of the vertical velocity, this is not seeding by the exact eigenfunctions of each mode to be excited.
Therefore to set up an eigenfunction there must be communication between different parts of the domain, which will happen by waves.
In the incompressible limit, the sound wave will effectively travel instantaneously.
The Alfv\'{e}n wave, however, will travel in a finite time, and the larger the wavelength along the magnetic field, the longer the timescale it will take for information to be communicated over that distance.
This may allow modes that may not have the largest growth rate to influence the structure because they start growing earlier and interchange modes, with constant phase along the magnetic field can take longer to develop.
This is likely to result in a wavelength dependence that is inversely proportional to $B$ (i.e. linear with the Alfv\'{e}n speed)  as well as the dependence proportional to $B^2$ in the growth rate.
In compressible simulations, the finite sound speed will also have to be considered which will influence both interchange and undular modes.}

\section{Energy partition of the instability}\label{en_part}

It is possible to give a definition for all the physical variables using the eigenfunction for $v_z$:
\begin{equation}
v_z(z)=\begin{cases}
v_z(0)\exp(kz), & \mbox{if } z<0\\
v_z(0)\exp(-kz),& \mbox{if } z>0
\end{cases}.
\end{equation}
Following the derivation in \citet{CHAN1961} the $x$ and $y$ components of the velocity field are defined in terms of $v_z$ as:
\begin{align}
v_x(z)=\frac{ik_x}{k^2}Dv_z= & \frac{ik_x}{k}\begin{cases}v_z, & \mbox{if } z<0 \\-v_z, & \mbox{if } z>0\end{cases}\\
v_y(z)=\frac{ik_y}{k^2}Dv_z= & \frac{ik_y}{k}\begin{cases}v_z, & \mbox{if } z<0 \\-v_z, & \mbox{if } z>0\end{cases},
\end{align}
where $D$ symbolises the derivative in the $z$ direction. 
For the magnetic field perturbations we have:
\begin{equation}
\sigma\mathbfit{b}=ik_yB\mathbfit{v} 
\end{equation}
which gives:
\begin{align}
b_x(z)=-\frac{k_xk_y}{k^2\sigma}B Dv_z(z)= & -\frac{k_xk_y}{k\sigma}B \begin{cases}v_z, & \mbox{if } z<0 \\-v_z, & \mbox{if } z>0\end{cases}\\
b_y(z)=-\frac{k_y^2}{k^2\sigma}B Dv_z(z)=- & \frac{k_y^2}{k\sigma}B \begin{cases}v_z, & \mbox{if } z<0 \\-v_z, & \mbox{if } z>0\end{cases}\\
b_z(z)= & i\frac{k_y}{\sigma}B v_z(z),
\end{align}
From this, it is possible to calculate the partition between kinetic and magnetic energies in the system.
Here the method presented in \citet{MOR2015} {for the determination of the energy of normal modes in linear MHD, which they applied to sausage modes,} will be adapted for this calculation.
{This method is applicable for a couple of simple reasons, a normal mode decomposition is used and averaging over one wavelength in those directions removes the influence of the non-perturbed component of the magnetic field, and that the energy of the mode is finite because the perturbed quantities are 0 as $z \rightarrow \pm \infty$ (or the $r$ direction for the sausage mode).
One change has been implemented, here the time integration is not performed as the temporal evolution is not cyclical.}
The equations for the energies as a function of $z$ are given as:
\begin{align}
KE(z)=\frac{1}{4}\rho_0\mathbfit{v}\cdot\mathbfit{v}^*\\
ME(z)=\frac{1}{16\pi}\mathbfit{b}\cdot\mathbfit{b}^*.
\end{align}
The $x$, $y$ and $t$ distributions are ignored as they add no important information to this discussion.
All that is left is to calculate $\mathbfit{v}\cdot\mathbfit{v}^*$ \& $\mathbfit{b}\cdot\mathbfit{b}^*$ and then integrate over $z$.

As laid out in the previous paragraph, next we calculate $\rho_0\mathbfit{v}\cdot\mathbfit{v}^*$ \& $\mathbfit{b}\cdot\mathbfit{b}^*$.
\begin{align}
\rho_0\mathbfit{v}\cdot\mathbfit{v}^*=\rho_0v_z^2\left(\frac{k_x^2}{k^2}+\frac{k_y^2}{k^2} +1 \right)=2\rho_0v_z^2\\
\mathbfit{b}\cdot\mathbfit{b}^*=v_z^2\frac{k_y^2B^2}{\sigma^2}\left(\frac{k_x^2}{k^2}+\frac{k_y^2}{k^2} +1 \right)=2v_z^2\frac{k_y^2B^2}{\sigma^2}
\end{align}
which gives $KE(z)$ and $ME(z)$ as:
\begin{align}
KE(z)=\frac{1}{4}\rho_0\mathbfit{v}\cdot\mathbfit{v}^*= & \frac{1}{2}\rho_0v_z^2\\
ME(z)=\frac{1}{16\pi}\mathbfit{b}\cdot\mathbfit{b}^*= & \frac{1}{8\pi}v_z^2\frac{k_y^2B^2}{\sigma^2}.
\end{align}
If both KE and ME are integrated with respect to $z$ (by integrating from $-\infty$ to the density discontinuity at $z=0$ and then from $z=0$ to $\infty$) we can calculate the total energies:
\begin{align}
TKE=& \int^{\infty}_{-\infty}\frac{1}{2}\rho_0v_z^2 dz \\= & \frac{1}{2}v_z(0)^2\left(\int^0_{-\infty}\rho_l\exp(2kz) dz +\int^{\infty}_0\rho_u\exp(-2kz) dz \right) \nonumber\\
= & \frac{1}{2}v_z(0)^2\frac{\rho_u+\rho_l}{2k}= \frac{1}{2}v_z(0)^2\frac{\rho_{av}}{k}\nonumber\\
TME=& \frac{1}{8\pi}\frac{k_y^2B^2}{\sigma^2}\int^{\infty}_{-\infty} v_z^2dz= \frac{1}{8\pi}\frac{k_y^2B^2}{k\sigma^2}v_z(0)^2,
\end{align}
where $\rho_{av}=(\rho_u+\rho_l)/2$.
It is clear that the TKE is only a function of $v_z(0)$ (the $z$ component of the velocity at $z=0$), the average density and the wavenumber, i.e. independent of the magnetic field.

The ratio of TME and TKE will give information about the relative amount of magnetic energy to kinetic energy.
\begin{equation}
TME/TKE=\frac{k_y^2B^2}{4\pi\rho_{av}\sigma^2}=\frac{\omega_A^2}{\sigma^2},
\end{equation}
where $\omega_A$ is the frequency of the Alfv\'{e}n wave calculated using the average density of the system.
From this ratio, it is easy to see that an interchange mode ($k_y=0$) gives $TME/TKE=0$ and an undular mode gives:
\begin{equation}
TME/TKE=\frac{\omega_A^2}{\left(Agk_y-\frac{k_y^2B^2}{2\pi(\rho_u+\rho_l)} \right)}=\frac{\omega_A^2}{Agk_y-\omega_A^2}.
\end{equation}
More interesting is the most unstable mode for a given $\theta$ where:
\begin{equation}
TME/TKE=1
\end{equation}
i.e. the most unstable mode for a given $\theta$ is the class of modes that equally partition the gravitational energy released by the instability into kinetic and magnetic energies.

\section{Nonlinear Saturation of the Linear Regime}\label{nonlin_regime}

Here we will think about the nonlinear saturation of a the magnetic Rayleigh-Taylor instability.
The first point to note is that the eigenfunction for the vertical velocity $\widetilde{v_z}$ is given by \citep{CHAN1961}:
\begin{equation}
\widetilde{v_z}(z)=v_z(0) \exp \left( - k |z|  \right)
\end{equation}
so the vertical scaling of the eigenfunction is $1/k$.
Therefore, there is an inherent implication that once the contact discontinuity has undergone a vertical deformation of lengthscale $\xi$ that is greater than $1/k$ then the instability can be seen to have deformed the interface to a greater extent than is given by the linear regime of the instability.
This implies that $1/k$ can be used as the vertical scale through which the discontinuity can be distorted before it enters its nonlinear evolution.

Dynamic arguments can also be used to produce this scaling.
Here we will use the linear relation:
\begin{equation}
\frac{\partial \xi}{\partial t}=\sigma \xi=v_z(0)
\end{equation}
One argument for the development of nonlinearity of the instability can be based on the nonlinear saturation through velocity shear driven instabilities like the Kelvin-Helmholtz instability.
If we balance the temporal derivative of the velocity with the advective derivative:
\begin{align}
\frac{\partial v}{\partial t}= & \sigma v \propto \mathbfit{v}\cdot\nabla v\sim kv^2\\
\implies &\frac{v}{\sigma}=\xi=\frac{1}{k}.
\end{align}
Therefore, it can be predicted that the Rayleigh-Taylor instability will develop secondary Kelvin-Helmholtz instabilities when the plumes have travelled through a distance of approximately $1/k$.

Another way of looking at this is that the nonlinear saturation can be understood as the point where there is a significant increase in magnetic forces to halt the further evolution of the rising/falling plumes.
\begin{align}
\sigma v =\frac{\partial v}{\partial t}=&\frac{1}{\rho_0}\frac{1}{4\pi}\mathbfit{j}\times\mathbfit{b} \sim \frac{1}{\rho_0}\frac{1}{4\pi}\frac{k_y^2k}{\sigma^2}B^2 v^2\\
\implies & \frac{v}{\sigma}=\xi\sim \frac{1}{k}\frac{\sigma^2}{\omega_A^2}.
\end{align}
This means that when the linear instability is dominated by the production of kinetic energy instead of the magnetic energy (see previous section), then the magnetic forces saturate at greater deformations of the boundary implying that the saturation dominates with the class of modes given by the most unstable mode for a given $\theta$ giving the equipartition of the two saturation mechanisms.
However, for modes where the magnetic energy dominates, the saturation will be for smaller deformations of the boundary.

\section{Thickening of the density jump by linear modes}\label{den_thick}

Here we look at the case where there is a sea of different plumes all excited by a random perturbation, the mode for a given $k$ will be the dominated by the modes that grow fastest for that given $k$, here for simplicity we can take system evolving under the whole spectrum of modes given by $\sigma^2=CAgk_u$ where $C$ is $0< C \le 1$ (NB: the following analysis deal with interchange modes only when $C=1$ or the spectrum of most-unstable-modes for a given $\theta$ when $C=1/2$), and lead to a thickening of the boundary between the two fluids over time.
If we just look at this thickening from the perspective of the growth of a linear mode and take its nonlinear saturation to occur when it reaches the height $1/k$ which then allows the next mode with smaller $k$ to take over until it reaches its own saturation.
The saturation height is given by the equation:
\begin{equation}
\xi_t=\xi_0\exp(\sigma_k t),
\end{equation}
where $\xi_t$ is the maximum vertical distance of the contact discontinuity from its initial position given by the magnitude of the perturbation $\xi_0$.
If we normalise the distance by some arbitrary wavenumber $\xi_t=\xi_t'k_0$ and the time by the growthrate of the preferred mode with this wavenumber then we have $t=t'/\sqrt{Agk_0C}=t'/\sigma_{k_0}$ the equation becomes:
\begin{equation}
\xi_t'=\xi_0'\exp(\sqrt{k'} t').
\end{equation}
If we know that at a time $t_0'$ the mode of wavenumber $k_0$ has saturated giving $\xi'_{t_0'}=1$ and that at time $t_1'$ the mode of wavenumber $k_1$ has saturated giving $\xi'_{t_1'}=1/k_1'$, the ratio of the separation given by these two saturated modes is:
\begin{equation}
\frac{1}{k_1'}=\exp(\sqrt{k_1'} t_1'-t_0').
\end{equation}
This can be rearranged to give:
\begin{equation}
t_1'=\frac{1}{\sqrt{k_1'}}\left[ t_0' +\ln \left(\frac{1}{k_1'} \right)  \right].
\end{equation}
From this we can expect the thickening of the boundary layer between the two fluids will increase due to the growth and saturation of linear modes as:
\begin{equation}
t \propto \sqrt{L} \ln (L),
\end{equation}
where $L$ is the thickness of the boundary.
This predicts that for large $L$ the boundary thickness will change as $L\sim t^2$.

Numerical experiments seem to suggest that the nonlinear thickening of the boundary layer  for the magnetic Rayleigh-Taylor instability follows the same $t \propto \sqrt{L}$ dependence \citep{STONE2007} as is seen in experiments of the hydrodynamic Rayleigh-Taylor instability \citep[e.g.][]{RIST2004}.
This suggests that in experiments or numerical simulations of the nonlinear evolution of this system, great care must be taken to make sure that the initial spectrum of perturbations for the instability is sufficiently narrowband {or that the system has been sufficient time to develop all possible linear modes} so that later evolution is dealing with {a thickness of the layer} far beyond the spectrum of perturbations and as such does not have any competition between larger scale linear modes and the nonlinear evolution.

\section{Influence of sheared magnetic field}

{The analysis provided so far in this paper has been concerned with a magnetic field that is in the same direction both above and below the discontinuity.
However, as has been seen in this work and countless others, this problem is ill posed because the growth rate is unbounded.
One way to make the problem well posed, as well as make it more applicable for many astrophysical situations, is to include the affect of magnetic shear.
The linear stability of this problem was investigated by, amongst others, \citet{RUD2014}.
The growth rate for a situation where the strength and direction of the magnetic field change across the discontinuity, but everything else stays the same as the situation under consideration here, is given in Equation 19 of \citet{RUD2014}.

\subsection{Basis vectors}

Looking at Equation 28 from \citet{RUD2014}, and with a change in notation to be consistent with this paper, the modes with the maximum and minimum growth rate for a fixed $k$ are given by the $\theta$ that satisfies this equation:
\begin{equation}
\tan \theta = \frac{(B_u/B_l)^2+\cos 2\alpha}{\sin 2\alpha}\pm\left[\frac{(B_u/B_l)^2+\cos 2\alpha}{\sin 2\alpha}+1 \right]^{1/2},
\end{equation}
where $\alpha$ is the angle between the magnetic fields above and below the discontinuity, $B_u$ is the field strength above the discontinuity and $B_l$ is the field stregnth below.
This implies that over the range of angles $-\pi/2 \le \theta \le \pi/2$ there is one maximum and one minimum in the growth rate, and as such the other two maximum and minimum that exist do not result in any more unique modes.
Therefore, for a given $k$, even in the presence of magnetic shear, there exist only two unique modes that have the same growth rate.
This means that the basis vector argument presented in Section \ref{BASIS} does not become redundant in the case of magnetic shear.

To illustrate this further, it is physically intuitive to look at the case where only the direction of the magnetic field changes across the boundary, but the strength remains the same.
In fact, the growth rate for the case where the strength of the magnetic field both above and below the discontinuity are the same and with choice of axis such that the parallel magnetic field is in the $y$ direction and the anti-parallel component (i.e. the component that reverses sign across the discontinuity) is in the $x$ direction gives:
\begin{equation}\label{shearRT}
\sigma^2=kg\left[A- \frac{B_x^2k_x^2+B_y^2k_y^2}{2 \pi (\rho_u + \rho _-)gk } \right ].
\end{equation}
This equation bears striking similarity to Equation \ref{RTeqn}, apart from the fact that it is clear that for no choice of wave vector $\mathbfit{k}$ does suppression from the magnetic field disappear.
For this case, the perturbations $[k_x,k_y]$ and $[-k_x,k_y]$ both give the same $\sigma$ and form a pair of basis vectors of $\rm I\!R^2$ of length $k$.
This could explain why though the width and length of the structure in the simulations of \citet{STONE2007} changes with magnetic shear it is still possible to explain the structure that develops through superposition.

\subsection{Most unstable modes}

The study of the most unstable modes of Equation \ref{shearRT} lends itself to the comparison with the case where there is no magnetic shear because of the similarity of the equations for the growth rate.
Slightly reworking Equation \ref{shearRT} gives:
\begin{equation}\label{shearRT1}
\sigma^2=kg\left[A- k\frac{B_x^2\sin^2\phi +B_y^2\cos^2\phi }{2 \pi (\rho_u + \rho _-)g } \right ],
\end{equation}
where $\phi$ is the angle between the wave vector and the $y$ axis.
Now we can look at the most unstable wave vector for a given $k$.
By taking the derivative with respect to $\phi$ of Equation \ref{shearRT1} and rearranging, we have:
\begin{equation}
(B_x^2-B_y^2)\sin 2\phi=0.
\end{equation}
In general, this implies that whenever the wavevector is in either the $x$ or $y$ directions, i.e. aligned either with the parallel or anti-parallel components of the magnetic field, depending on which is weakest.
That is to say, the fastest growing mode for a given $k$ is, unsurprisingly, the one which does least work against the magnetic field.
However, in the case where $B_x=B_y$, i.e. the magnetic field above the discontinuity is at an angle of $\pi/2$ to the magnetic field below, when any angle satisfies the condition.
In this case the instability will form plumes of circular cross-section similar by the superposition of two wave vectors at $\pi/2$ to each other similar to those found in the hydrodynamic Rayleigh-Taylor instability or the simulations with large magnetic shear of \citet{STONE2007}.
It is also worth noting that, as with the undular mode of the magnetic Rayleigh-Taylor instability, for sufficiently small $k$, the growth rate will scale as $\sigma\propto k^{1/2}$.

Now we can look at the most unstable wavenumber for a given angle $\phi$.
By taking the derivative with respect to $k$ of Equation \ref{shearRT1} and rearranging, we have:
\begin{equation}
k_u=\frac{g \pi (\rho_u-\rho_l)}{\sin^2\phi B_x^2 + \cos^2\phi B_y^2}.
\end{equation}
This is now bounded both above and below by:
\begin{equation}
\frac{g \pi (\rho_u-\rho_l)}{B_y^2}\le k_u\le\frac{g \pi (\rho_u-\rho_l)}{B_x^2},
\end{equation}
assuming that $B_x>B_y$.
There clearly exists a range of solutions for $k$ resulting for the varying of $\phi$ for $\theta \in [0,\pi/2]$.
Therefore, we can define the set of growthrates that correspond with these most unstable modes:
\begin{equation}\label{most_growth}
\sigma^2=\frac{1}{2}A k_u g.
\end{equation}
This is the same spectra as found for the uniform magnetic field case.
It can be hypotesized from here that a class of modes that satisfy:
\begin{equation}
\sigma^2=CA k_u g.
\end{equation}
for any $C$ such that $0\le C\le 1/2$.

In this system, a most unstable mode for the whole system can be defined as $\sigma^2=1/2 A k_u g$ for 
\begin{equation}
k_u=\frac{g \pi (\rho_u-\rho_l)}{B_y^2}.
\end{equation}
assuming $B_x>B_y$.
Again it is clear that the most unstable mode for this system is one without a pair, and so this again raises the question as to why these modes are not seen in the simulations of \citet{STONE2007}.

As the $\sigma\propto k^{1/2}$ dependence is also present in this case for sufficiently small $k$ and because the physics discussed has not changed, it can be expected that the development of non-linearities follows the same restrictions as for the non-sheared case (see Section \ref{nonlin_regime}).
However, in this case the magnetic saturation has a factor $k_x^2B_x^2+k_y^2B_y^2$ instead of $(k_yB)^2$.
From this it is clear to see that the arguments presented for the thickening of the mixing layer through linear modes (see Section \ref{den_thick}) will not change as a result of the introduction of magnetic shear.

\subsection{Energy partition for a sheared magnetic field}

To calculate the energy partition of the instability in the sheared field case, it is simplest to first look at the case relating to Equation \ref{shearRT}.
As we are going to integrate from $-\infty$ to the boundary at $z=0$ and then from the boundary to $\infty$, the existence of a delta function in the current does not cause any problems for the analysis.
In fact it means that other than the existence of an $x$ component of the magnetic field and that it has a different sign above and below the discontinuity, the analysis is practically the same as Section \ref{en_part}, with the eigenfunction for $v_z(z)$ and kinetic energy not changing at all.
The magnetic field perturbations are given as
\begin{align}
b_x(z)=&-\frac{k_x}{k^2\sigma}(\mathbfit{k} \cdot \mathbfit{B}) Dv_z(z)\\=& \frac{k_x}{k\sigma}\begin{cases}(-k_yB_y+k_xB_x) v_z, & \mbox{if } z<0 \\(k_yB_y+k_xB_x)v_z, & \mbox{if } z>0\end{cases}\nonumber\\
b_y(z)=&-\frac{k_y}{k^2\sigma}(\mathbfit{k} \cdot \mathbfit{B})Dv_z(z)\\ = & \frac{k_y}{k\sigma}\begin{cases}(-k_yB_y+k_xB_x) v_z, & \mbox{if } z<0 \\(k_yB_y+k_xB_x)v_z, & \mbox{if } z>0\end{cases}\nonumber\\
b_z(z)= & i\frac{\mathbfit{k}\cdot \mathbfit{B}}{\sigma} v_z(z)=\frac{i}{\sigma}\begin{cases}(-k_yB_y+k_xB_x) v_z, & \mbox{if } z<0 \\(k_yB_y+k_xB_x)v_z, & \mbox{if } z>0\end{cases}.
\end{align}

Performing the same steps as Section \ref{en_part} gives 
\begin{align}
TKE= & \frac{1}{2}v_z(0)^2\frac{\rho_{av}}{k}\\
TME=& \frac{1}{8\pi}\frac{k_x^2B_x^2+k_y^2B_y^2}{k\sigma^2}v_z(0)^2\label{TME_shear}\\
TME/TKE=&\frac{k_x^2B_x^2+k_y^2B_y^2}{4\pi\rho_{av}\sigma^2}=\frac{\omega_{Ax}^2+\omega_{Ay}^2}{\sigma^2},
\end{align}
where $\omega_{Ax}$ and $\omega_{Ay}$ are the Alfv\'{e}n frequencies calculated from the magnitudes of the $x$ and $y$ components of the magnetic field respectively.
One clear difference to the case with a uni-directional magnetic field is {that there is no longer} a perturbation that does not increase the magnetic energy of the system.
Again, the case where the ratio is $1$ corresponds to the set of most unstable modes for a given $\theta$.

Referencing back to the more general case where the magnetic field is allowed to have both different strengths and directions across the discontinuity, by comparison with Equation \ref{TME_shear} we can see that the total magnetic energy is given by:
\begin{equation}
TME= \frac{1}{16\pi}\frac{(\mathbfit{k}\cdot \mathbfit{B})^2_l+(\mathbfit{k}\cdot \mathbfit{B})^2_u}{k\sigma^2}v_z(0)^2.
\end{equation}
}

\section{Discussion}

To conclude the main points presented.
For the linear growth of the magnetic Rayleigh-Taylor instability where perturbations are both "long" in that the ratio of the wavelength to thickness of the discontinuity is large ($\lambda/\Delta x >> 1$), and "short" in that both the ratio of the wavelength to the gas pressure scale height is small ($\lambda/\Lambda_g < 1$) and the ratio of the wavelength to the characteristic height over which the magnetic field changes its direction is small ($\lambda/L_B < 1$), I have determined a set of key characteristics from the growth of the instability {for the case where the magnetic field is uniform and with a shear at the contact discontinuity}.
\begin{itemize}

\item In a 3D system, all perturbations that have both a component parallel ($\lambda_{y}$) and perpendicular ($\lambda_x$) to the direction of the magnetic field have a perturbation pair of $\lambda_x$ and $-\lambda_y$ giving a set of basis vectors for $\rm I\!R^2$ that has the same growth rate. 
It is the superposition of these two perturbations that leads to an interference pattern giving filamentary structure aligned with the magnetic field.

%\item Because they do not have a perturbation pair, the pure interchange and undular modes must have a direction of constant phase and therefore a correlation velocity that is infinite in that direction. Therefore, these modes cannot grow from a random perturbation in a 3D system that has finite velocities associated with the transport of information.

\item For the set of most unstable modes for a given $\theta$ the wavelength along the magnetic field tends to zero as the wavelength across the magnetic field tends to zero. However, the aspect ratio between the two scales increases as $k_x \propto k_y^2$.

\item It cannot be said that the magnetic field suppresses small scales (in the sense that small scales along the magnetic field cannot exist), as I show that the equation predicts that the fluctuations of the magnetic field can be produced down to arbitrarily small scales. The magnetic fields key role is in determining the aspect ratio between the wavelength parallel and perpendicular to the magnetic field

\item The growth rate of the instability for the set of most unstable modes for a given $\theta$ is given by: $\sigma^2=1/2A g k_{u}$, which is independent of magnetic field strength. Those for a given $k$ are given by: $\sigma^2=A g k_{u}$

\item The energy partition between kinetic and magnetic energy for the release gravitational energy is such that equal energy is given to kinetic and magnetic energy

\item The nonlinear stage of the instability begins once the boundary between the two fluids has deformed in the $z$ direction by approximately a distance of $1/k$.  Though for modes where magnetic energy dominates kinetic energy then this happens for: $\omega_A^2/(\sigma^2k)$.

\item When looking at the temporal evolution of the density discontinuity as a result of subsequent linear modes from the set of most unstable modes (either for a given $\theta$ or $k$), at later times the linear modes will result in the boundary layer ($L$) thickening as $L\propto t^2$.

\item {The results hold in the more general case where there is shear in the magnetic field at the contact discontinuity.}
\end{itemize}

One of the first uses for this knowledge on the linear growthrate of the magnetic Rayleigh-Taylor instability is its application to the inverse problem of determining the magnetic field strength from astrophysical objects.
By taking observed wavelengths, these can be related to magnetic field strengths by assuming that the system is growing under a most unstable mode.
However, the physics of the system do not prefer any particular wavelength, in fact the system wants $k \rightarrow \infty$ implying that if a clear wavelength is observed for a system, i.e. the observations of the Rayleigh-Taylor instability in prominences by \citet{BERG2011}, then either a narrowband driver has been applied to the perturbations or the physical assumption that went into deriving Equation \ref{RTeqn} are invalid.
The other issue is that the equation for the most unstable mode gives a pair of wavelengths perpendicular and parallel to the magnetic field ($k_x$ and $k_y$) that are connected by an aspect ratio that is a function of the magnetic field strength.
Therefore, if you have not measure both $k_x$ and $k_y$ then there are an infinite rage of solutions that give a fixed wavelength for an infinite range of magnetic field strengths.

A key point that needs to be understood is that the instability when excited by a random perturbation that contains a broad spectrum of modes becomes fundamentally a 3D system and cannot be truly captured by 2D simulations.
Unlike the hydrodynamic case, the MHD evolution is no longer isotropic due to the addition of the magnetic field, and therefore a 2D simulation would only be able to capture the growth of a single mode from the whole spectrum of preferred modes.
It is worth noting that there is a fundamental issue with 3D simulations as well.
In 3D simulations, often periodic boundaries are introduced to remove the necessity for a domain of infinite size, this however has its own drawbacks.
Given a simulation domain of horizontal size $L_x$ by $L_y$, modes that satisfy the following relation $k_xN_x=L_x$ and $k_yN_y=L_y$ where $N_x$,\,$N_y \in \rm I\!N$ are resonant with the box size and preferentially form.

{One caveat to this estimate is that at later times, physically it can be expected that the modes are not growing from a discontinuity, but from a layer of ever increasing thickness.
While all modes are linear then the analysis presented holds, but there will be a nonlinear feedback as modes saturate which has not been modelled here.
One aspect where this feedback will become important is with the creation of a finite region where the density transitions, especially as this will allow instability modes that are on scales smaller than the thickness of the layer to form at multiple heights throughout the layer.
However, it can be assumed that for a mode that when it begins to grow has a wavelength that is much bigger than the thickness of the layer can be initially approximated by the same mode growing from a discontinuity and it is only as it approaches its saturation that the layer thickness will become comparable with the wavelength of the mode, so roughly speaking a linear approximation from a discontinuity can be expected to approximate the growth of this mode.
A full analysis of these effects is beyond the scope of this paper. }

We should also take into account the equation for the growth rate of the most unstable mode, Equation \ref{mum}.
This equation states that for a given most unstable mode the larger the wavenumber of this unstable mode the larger the growth rate.
Therefore, for a truly random perturbation, the mode with the largest possible wavenumber would grow quickest.
This implies that the physics of the system is heavily biased towards creating structures at the largest possible wavenumber{, i.e. the growth rate is unbounded with respect to $k$,} and so the large wavelength plumes observed in both quiescent prominences \citep{BERG2011} and in the infalling material presented in \citet{INN2012,CAR2014} cannot be explained as the growth of the most unstable mode of the simple system under study here.
Therefore, if relatively large wavelengths are the most unstable, then there must be more complexity in the system, e.g. a finite width to the density transition layer or shear in the magnetic field \citep{RUD2014}, which will necessarily change the conditions for most unstable mode.
{In fact, a more suitable equation to use would be Equation \ref{shearRT} as long as there exists a physical justification, e.g. both fluids are low $\beta$, for taking the strength above and below the discontinuity as being the same.
This would justify the statement in \citet{CAR2014} that the field strength found when assuming a constant magnetic field can be taken as a lower limit for the field strength because of the potential existence of magnetic shear.}

\section*{Acknowledgements}

AH would like to thank Drs Adrian Barker, Takeshi Matsumoto and Jack Carlyle for their comments on the manuscript which greatly improved the content of the manuscript.
AH would like to thank the anonymous referees for their comments which greatly improved the quality of the manuscript. 
AH is supported by his STFC Ernest Rutherford Fellowship grant number ST/L00397X/1.

%%%%%%%%%%%%%%%%%%%%%%%%%%%%%%%%%%%%%%%%%%%%%%%%%%



%\bibliographystyle{mnras}


%%%%%%%%%%%%%%%%%%%%%%%%%%%%%%%%%%%%%%%%%%%%%%%%%%

% Don't change these lines
\bsp	% typesetting comment
\label{lastpage}
\end{document}